\documentclass[12pt]{article}
\usepackage{latexsym,graphicx}
\usepackage{subfigure}
\usepackage{amssymb}
\usepackage{amsmath}
\usepackage{amscd}
\usepackage{amsthm}
\usepackage{float}
\usepackage[left=2cm,top=2.5cm,right=2.5cm,bottom=1.5cm]{geometry}    
\begin{document}
\begin{center}
\large{\bf{A Cyclic Universe With Varying Cosmological Constant in $f(R,T)$} gravity} \\
\vspace{10mm}
\normalsize{Nasr Ahmed$^1$ and Sultan Z. Alamri$^2$}\\
\vspace{5mm}
\small{\footnotesize $^1$ Astronomy Department, National Research Institute of Astronomy and Geophysics, Helwan, Cairo, Egypt\footnote{nasr.ahmed@nriag.sci.eg}} \\
\small{\footnotesize $^2$ Mathematics Department, Faculty of Science, Taibah University, Saudi Arabia\footnote{szamri@taibahu.edu.sa}} \\
\end{center}  
\date{}
\begin{abstract}
A new kind of evolution for cyclic models in which the Hubble parameter oscillates and keeps positive has been explored in a specific $f(R,T)$ gravity reconstruction. A singularity-free cyclic universe with negative varying cosmological constant has been obtained which supports the role suggested for negative $\Lambda$ in stopping the eternal acceleration. The cosmological solutions have been obtained for the case of a flat universe,  supported by observations. The cosmic pressure grows without singular values, it is positive during the early-time decelerated expansion and negative during the late-time accelerating epoch. The time varying EoS parameter $\omega(t)$ shows a quintom behavior and is restricted to the range $-2.25 \leq \omega(t) \lesssim \frac{1}{3}$. The validity of the classical linear energy conditions and the sound speed causality condition has been studied. The non-conventional mechanism of negative cosmological constant that are expected to address the late-time acceleration has been discussed. 

\end{abstract}
PACS: 04.50.-h, 98.80.-k.\\
Keywords: Modified gravity, Cosmology, Cyclic universe.
\section{Introduction and motivation}

Recent observations indicate that the universe is highly homogeneous and isotropic on large
scales \cite{teg,ben,sp}, and expanding at an accelerating rate \cite{11,13,14}. A hypothetical form of energy with
negative pressure known as dark energy (DE) has been assumed as the reason behind this cosmic acceleration. Several dynamical scalar fields DE models have been proposed to explain this late-time acceleration including quintessence \cite{quint}, Chaplygin
gas \cite{chap}, phantom energy \cite{phant}, k-essence \cite{ess}, tachyon \cite{tak} and ghost condensate \cite{ark,nass}. Modified gravity represents another approach to explain the accelerating expansion based on modifying the geometrical part of the Einstein-Hilbert action \cite{moddd}, it has been successful in explaining the galactic rotation curves without the need to dark matter assumption \cite{noj1, de1}. Examples include $f(R)$ gravity \cite{39} which is widely regarded as the most popular modified gravity theory where $R$ is the Ricci scalar, Gauss-Bonnet gravity \cite{noj8} and f(T) gravity \cite{torsion} where $T$ is the torsion scalar. While $f(R)$ gravity modifies Einstein equations by replacing curvature scalar $R$ in the Einstein-Hilbert action by an arbitrary function $f(R)$, the torsion scalar $T$ is used as a fundamental ingredient in $f(T)$ gravity instead of $R$. Moreover, $f(T)$ gravity doesn't have the local Lorentz symmetry in general which generates some difficulties as has been illustrated in \cite{tor2}. Driven by the torsion effects, $f(T)$ gravity can explain the accelerated expansion without assuming a new form of energy \cite{beng1}. In Gauss-Bonnet gravity, the Einstein-Hilbert action is modified by replacing $R$ by the the Gauss-Bonnet term $G=R^2-4R^{\mu\nu}R_{\mu\nu}+R^{\mu\nu\rho\delta}R_{\mu\nu\rho\delta}$. More generally, a function $f(G)$ can be used. $f(R)$ gravity has been generalized by introducing an arbitrary function $f(R,T)$ \cite{1}, where $T$ is the trace of the energy momentum tensor. Some cosmological aspects of $f(R,T)$ gravity have been investigated in \cite{46,47,48}.\par

Some different alternatives to dark energy have recently been proposed \cite{alt1, alt2, alt3}. In \cite{alt1}, it has been shown that a classical gravity theory with massive gravitons can lead to a consistent cosmological model that fits the observational data without dark energy assumption. The hypothesis of the scale invariance of the macroscopic empty space has been used in \cite{alt2} and led to cosmic acceleration and Keplerian evolution of the galactic rotation curve without any need to dark energy or dark matter. The bulk viscosity has been suggested as an alternative to DE in \cite{alt3}. It has been found that for a sufficiently large bulk viscosity, the effective pressure becomes negative and could mimic a DE equation of state.   \par

In the current work, we use a specific reconstruction of $f(R,T)$ gravity with varying cosmological constant. Varying cosmological constant has been introduced in the literature as an attempt to solve some of the standard $\Lambda CDM$ model problems such as the fine-tuning and cosmic coincidence. Several cosmological scenarios with variable cosmological constant have been proposed \cite{shap, bona, mavro, perrr, soco, sta, nasa}. Since a very small positive energy with enough negative pressure can act as a repulsive gravity that accelerates the cosmic expansion, the cosmological constant can be considered as the simplest candidate for dark energy. The possibility for cosmic acceleration with negative cosmological constant has also been discussed in the literature where negative cosmological constant can halt eternal acceleration \cite{vinc}. In this paper, we consider this last scenario of negative cosmological constant in an oscillating universe with very small positive energy density and enough negative pressure.\par

The cosmological constant represents the energy density of the vacuum. Although observations suggest a very small positive $\Lambda$ with magnitude $\approx 10^{-123}$ \cite{11,76,77}, it has been shown that a negative $\Lambda$ which can fit quite well a large data set is also possible and can provide a solution to the 'eternal acceleration' problem \cite{vinc}. An accelerating universe doesn't have to keep accelerating forever and it has been shown that such eternal acceleration is a consequence of assuming a positive $\Lambda$ to explain the cosmic accelerated expansion \cite{vinc}. This radical different approach of negative cosmological constant has been investigated by many authors \cite{vinc, gardena, grand, gu, prok, land, ma, ba, ch}. A strong argument for the negative cosmological constant has been presented by the AdS/CFT correspondence \cite{ah}. In 2011, Prokopec introduced modifications to Friedmann cosmology that can lead to observationally viable cosmologies with $\Lambda<0$ \cite{prok}. Maeda and Ohta \cite{ma} investigated gravitational theories with the Gauss-Bonnet curvature squared term and found a stable de Sitter solution with negative $\Lambda$. McVittie solution with $\Lambda<0$ has been studied by Landry et al in 2012 and it has been found that a negative $\Lambda$ ensures collapse to a Big Crunch as in pure FRW case \cite{land}.  Chruściel et al \cite{ch} proved the existence of large families of solutions of Einstein-complex scalar field equations with $\Lambda<0$. Recently, the variable cosmological constant has been given a new interpretation as thermodynamic pressure in the context of black hole physics both in AdS and dS background \cite{BHch}. As has been indicated in \cite{BHch}, This interpretation naturally follows from the realization that negative $\Lambda$ induces a positive vacuum pressure in space-time. A cyclic universe with negative cosmological constant has been studied in \cite{bis}. The idea of associating $\Lambda$ with pressure was investigated in \cite{chem1, chem2} from several perspectives. So, the negative varying cosmological constant in the current work has a strong base in the literature and is supported by results of other researchers.\par

One of the major problems in the standard big bang cosmology is the initial singularity. A possible way to solve this problem is to build an oscillatory model where the current universe is as a result of the collapse of a previous universe.
Constructing a consistent oscillating cosmological models free from initial singularity is an old problem in theoretical cosmology. The second law of thermodynamics implies that entropy increases from cycle to cycle which means that successive cycles become longer and larger. This situation leads to initial singularity if we keep extrapolating back in time (as the cycle duration converges to zero) and then removes the motivation to study such oscillating models \cite{c1,c2,c3,c4}. Moreover, the oscillating universe in these old models is finite and closed with zero cosmological constant while recent observations favor a flat universe with non-zero cosmological constant. The singularity theorems \cite{hawk} proves that a modification or replacement of General Relativity is required to avoid cosmic singularity. The idea of an infinitely oscillatory universe remained abandoned until the discovery of dark energy in 1998 when people started to make use of DE to overcome the entropy problem. cyclic cosmological models that exploit DE and can generate homogeneity, flatness and density fluctuations without the need for inflation have been constructed in \cite{c5,c6,c8,c9}.  While the inflationary model is in a very good agreement with observations \cite{marv}, answers to some basic questions regarding inflation still missing such as why did the universe start to inflate, What is the inflaton and why are the inflaton interactions finely-tuned? A cyclic Model inspired by M theory, in which the universe consists of two branes separated by a bulk, has been suggested in \cite{c6} as an alternative to the standard big bang/inflationary model. In this model, matter and energy densities don't diverge at the bounce (colliding of the two branes) and the entropy doesn't build up from cycle to cycle. Because the equation of state parameter has the condition $\omega \gg 1$ during the contraction phase, the universe is homogeneous, isotropic and flat. A different cyclic model solves the entropy problem has been suggested in \cite{pro} with $\omega <-1$ (phantom energy) throughout a cycle. It has also been shown in \cite{dod} that oscillating models can resolve the coincidence problem due to the natural variation of the periods of acceleration. The possibility for oscillating universe with quintom matter has been studied in \cite{os1}. The evolution of oscillating universe in massive gravity has been investigated in \cite{os2}.\par

In \cite{basic}, a 4-dimensional FRW oscillating cosmological model with quintom matter has been constructed through the following proposed time periodic deceleration parameter
\begin{equation} \label{q}
q(t)=-\frac{\ddot{a}a}{\dot{a}^2}=m \cos kt -1, 
\end{equation}
where $m$ and $k$ are positive constants. Equation (\ref{q}) leads to the following scale factor
\begin{equation} \label{a}
a(t)=A \exp \left[ \frac{2}{\sqrt{c^2-m^2}}\arctan \left(\frac{c\tan \left(\frac{kt}{2}\right)+m}{\sqrt{c^2-m^2}}\right)\right]
\end{equation}
where $A$ and $c$ are integration constants. In order to avoid singularities such as Big Rip, we need to choose $c>m$ \cite{basic}. In section (\ref{sol}) we show that in addition to the condition $c>m$, the difference $c-m$ must be very small to get a good agreement with observations. The scale factor (\ref{a}) provides a new kind of oscillating behavior where the Hubble parameter $H=\frac{\dot{a}}{a}$ oscillates but remains positive (In bouncing cosmology, the Hubble parameter $H$ passes from $H < 0$ to $H > 0$ and vanishes at the bounce point \cite{ali}). Because there are no singularities, the cyclic scenario represented by (\ref{a}) offers a possibility for the unification of the late-time acceleration and inflation where the universe passes through the inflationary era periodically without singularities. For a universe described by FRW metric, the relation between the scale factor $a$ and the redshift $z$ is $a=\frac{1}{1+z}$. Using this relation along with (\ref{a}), we can express the cosmic time in terms of $z$ as
\begin{equation} \label{t}
t=\frac{2}{k} \arctan \left[\frac{1}{c}\left(\sqrt{c^2-m^2}-m\right)\tan \left(\frac{1}{2}(\sqrt{c^2-m^2}) \ln\left(\frac{1}{A(1+z)}\right)\right)\right]
\end{equation}
The paper is organized as follows: section 1 is an introduction. In section 2, we derive the modified field equations with variable cosmological constant from $f(R,T)$ gravity action. In section 3, we make use of the time periodic deceleration parameter ansatz to generate oscillating solutions, then we provide a detailed discussion for the solutions. The stability analysis of the model is discussed in section 4. The final conclusion is included in section 5. 

\section{Field equations} 
The $f(R,T)$ gravity action is given by \cite{1}
\begin{equation}
S=\frac{1}{16\pi}\int{f(R,T)\sqrt{-g}d^{4}x}+\int{L_{m} \sqrt{-g}d^{4}x} , 
\end{equation}
where $L_{m}$ is the matter Lagrangian density. By varying the action $S$ with respect to $g^{\mu \nu}$, we obtain the field equations of $f(R,T)$ gravity as 
\begin{equation} \label{FieldEquations}
f_{R}(R,T)R_{\mu \nu}-\frac{1}{2} f(R,T)g_{\mu \nu}+(g_{\mu \nu} \Box  -\nabla_{\mu} \nabla_{\nu})f_{R}(R,T) 
=8\pi T_{\mu \nu}-f_{T}(R,T)T_{\mu \nu}-f_{T}(R,T)\Theta_{\mu\nu}.
\end{equation}
where $\Box = \nabla^{\mu}\nabla_{\mu}$, $f_{R}(R,T)=\frac{\partial f(R,T)}{\partial R}$, $f_{T}(R,T)=\frac{\partial f(R,T)}{\partial T}$ and $\nabla_{\mu}$ denotes the covariant derivative. $\Theta_{\mu \nu}$ and the stress-energy tensor $T_{\mu \nu}$ are given by
\begin{equation}
\Theta_{\mu \nu}=-2T_{\mu\nu}-pg_{\mu\nu}, ~~~~T_{\mu\nu}=(\rho+p)u_{\mu}u_{\nu}-pg_{\mu\nu}.
\end{equation}
The four-velocity $u_{\mu}$ satisfies $u_{\mu}u^{\mu}=1$ and $u^{\mu}\nabla_{\nu}u_{\mu}=0$. $\rho$ and $p$ are the energy density and pressure of the fluid respectively. The divergence of the energy-momentum tensor is given by \cite{correction}
\begin{equation} \label{non}
\nabla^{\mu} T_{\mu\nu}=\frac{f_T(R,T)}{8 \pi-f_T(R,T)}[(T_{\mu \nu}+\Theta_{\mu \nu})\nabla^{\mu}\ln f_T(R,T)+\nabla^{\mu}\Theta_{\mu\nu}-\frac{1}{2}g_{\mu\nu}\nabla^{\mu} T].
\end{equation}
Which represents the violation of energy-momentum conservation in $f(R,T)$ gravity. Different theoretical models can be obtained for each choice of the function $f(R,T)$. Taking $f(R,T)=f_{1}(R)+f_{2}(T)$, the gravitational field equations (\ref{FieldEquations}) becomes
\begin{equation}\label{Field Equations2}
f^{'}_{1}(R)R_{\mu \nu}-\frac{1}{2} f_{1}(R)g_{\mu \nu}+(g_{\mu \nu} \Box  -\nabla_{\mu} \nabla_{\nu})f^{'}_{1}(R)= 
8\pi T_{\mu \nu}+f^{'}_{2}(T)T_{\mu \nu} +\left(f^{'}_{2}(T)p+\frac{1}{2}f_{2}(T)\right) g_{\mu\nu}. 
\end{equation}
We simply take $f_{1}(R)=R$ and $f_{2}(T)=T$, so $f(R,T)= (R+T)$. In the case $f_2(T)\equiv 0$, we re-obtain the field equations of the standard general relativity. Now, equation (\ref{Field Equations2}) becomes 
\begin{equation}
R_{\mu \nu}-\frac{1}{2} g_{\mu \nu}R = 8\pi T_{\mu \nu}+ T_{\mu \nu}+ (P+\frac{1}{2}T)g_{\mu \nu}. \label{Fsubst}
\end{equation}
Which could be rearranged as
\begin{equation}
G_{\mu\nu}-\left(p +\frac{1}{2} T\right) g_{\mu\nu}=(8\pi+1) T_{\mu \nu}.    \label{ours}
\end{equation}
Where $G_{\mu\nu}=R_{\mu\nu}-\frac{1}{2} g_{\mu\nu} R$ is the Einstein tensor. Recalling Einstein equations with cosmological constant
\begin{equation}
G_{\mu\nu}+\Lambda g_{\mu\nu}=8\pi T_{\mu \nu}. \label{Einst}
\end{equation}
The term $\left(p +\frac{1}{2} T\right)$ can now be considered as the negative of the cosmological constant. i.e, $p +\frac{1}{2} T \equiv -\Lambda$. We could have easily used the form $f(R,T)=R+2\lambda T$ proposed by Harko et al in \cite{1} which has been extensively used in the literature, but our choice here leads to a specific formula for the varying cosmological constant. In \cite{sah}, the time periodic deceleration parameter (\ref{q}) has been used with the choice $f(R,T)=R+2\lambda T$ and no cosmological constant, it has been found that this leads to singularities in the cyclic behavior of $p$, $\rho$, $H$ and $a$. For the perfect fluid energy-momentum tensor we have $T^1_1=T^2_2=T^3_3=-p(t)$ and $T^4_4=\rho(t)$ and then the trace $T=\rho(t)-3p(t)$. So, (\ref{ours}) can be written as  
\begin{equation} \label{ours2}
G_{\mu\nu}-\frac{1}{2}\left(\rho-p\right) g_{\mu\nu}=(8\pi+1) T_{\mu \nu}.   
\end{equation}
And we have the following expression for the varying cosmological constant:
\begin{equation}
\Lambda=-\frac{1}{2}(\rho-p) \label{cosm},
\end{equation} 
Which also represents the energy density of the vacuum. The field equations (\ref{ours2}) is an improved form of the equations derived for the first time in \cite{46} using $f(R,T)=\lambda(R+T)$ (and used later by many authors). The choice $f(R,T)=\lambda(R+T)$ leads to a prefactor $\lambda$ in the Ricci scalar term which is not favored observationally at least on the solar system scale where GR is very successful, and a fine-tuning becomes necessary. The expression ($\ref{cosm}$) for varying $\Lambda$ is nothing but the negative of the thermodynamical work density \cite{work1,work2} 
\begin{equation}\label{int}
W=\frac{1}{2}(\rho-p)
\end{equation} 
where $\rho$ and $p$ are the energy density and pressure of cosmic matter, i.e. $\Lambda(t)=-W(t)$. It has been shown that the FRW cosmological equations can be expressed as $dE = T dS + W dV$ at the apparent horizon where $E = \rho V$ is the total energy and $W=\frac{1}{2}(\rho-p)$ is the work density \cite{work1,work2}. In the current $f(R,T)$ gravity model, the relation $\Lambda(t)=-W(t)$ allows a thermodynamical interpretation to the varying $\Lambda$ . Another thermodynamical interpretation has been given to $\Lambda$ in the framework of black hole physics as thermodynamic pressure \cite{BHch}. The generalization of the laws of black hole mechanics with non-zero $\Lambda$ leads to the following generalized first law of black hole thermodynamics \cite{BHch,bh2}
\begin{equation}
\delta M= T \delta S +V \delta P +\Omega \delta J + \Phi \delta Q
\end{equation} 
Where 
\begin{equation}
P=-\frac{\Lambda}{8\pi}
\end{equation} 
is interpreted as thermodynamic pressure. While the thermodynamic interpretation in \cite{BHch} is a natural consequence of negative $\Lambda$ which induces a positive vacuum pressure, the thermodynamic interpretation of (\ref{cosm}) arises dynamically from the field equations.
The non-conservation of energy-momentum tensor (\ref{non}) in the current $f(R,T)$ gravity reconstruction is given by 
\begin{equation} 
\nabla^{\mu} T_{\mu\nu}=-\frac{1}{8\pi+1}\left[\nabla^{\mu} (pg_{\mu\nu})+\frac{1}{2}g_{\mu\nu}\nabla^{\mu}T\right].
\end{equation}
As has been mentioned in \cite{sah}, The role of energy-momentum non-conservation in modified gravity have yet not been investigated properly in literature. For example, it has been shown that such violation of energy momentum conservation in modified gravity leads to accelerated expansion \cite{jos}. The FRW metric is given by
\begin{equation}
ds^{2}=-dt^{2}+a^{2}(t)\left[ \frac{dr^{2}}{1-Kr^2}+r^2d\theta^2+r^2\sin^2\theta d\phi^2 \right] \label{RW}
\end{equation} 
where $K$ is either $0$, $-1$ or $+1$ for flat, open and closed universe respectively. Applying equation (\ref{ours2}) to the metric (\ref{RW}) and taking (\ref{cosm}) into account, we get the following cosmological equations
\begin{eqnarray}
\frac{\dot{a}^2+K}{a^2} &=& \frac{(8\pi+1)\rho+\Lambda(t)}{3}.  \label{RW1}\\
\frac{\ddot{a}}{a} &=&  -\frac{(8\pi+1)}{6} (\rho+3p)+\frac{\Lambda(t)}{3}. \label{RW2}
\end{eqnarray}
Where $\Lambda(t)$ is given by (\ref{cosm}). Equations (\ref{RW1}) and (\ref{RW2}) are two differential equations in three unknown functions $a(t)$, $p(t)$ and $\rho(t)$. In the following section we discuss the exact solution.

\section{Solutions}\label{sol}

Solving (\ref{RW1}) and (\ref{RW2}) for $K=0$, the case supported by observations \cite{ben,sp,flat}, and making use of (\ref{a}) we get the  following expressions for the pressure $p(t)$, energy density $\rho(t)$, cosmological constant $\Lambda(t)$ and EoS parameter $\omega(t)$ (the expression for $q(t)$ is expressed by equation (\ref{q}))

\begin{equation}
p(t)=\frac{1}{64}\frac{32mk^2(\pi+\frac{1}{16})\cos^2(\frac{1}{2}kt)-k^2(16\pi m+24\pi+m+3)}{(\pi^2+\frac{\pi}{8})\left(-4m^2\cos^4(\frac{1}{2}kt)+4m^2\cos^2(\frac{1}{2}kt)+4cm\cos(\frac{1}{2}kt)\sin(\frac{1}{2}kt)+c^2\right)}
\end{equation}

\begin{equation}
\rho(t)=-\frac{1}{64}\frac{k^2\left(2m\cos^2(\frac{1}{2}kt)-24\pi-m-3\right)}{(\pi^2+\frac{\pi}{8})\left(-4m^2\cos^4(\frac{1}{2}kt)+4m^2\cos^2(\frac{1}{2}kt)+4cm\cos(\frac{1}{2}kt)\sin(\frac{1}{2}kt)+c^2\right)}
\end{equation}

\begin{equation}
\Lambda(t)=\frac{1}{8}\frac{k^2\left(2m\cos^2(\frac{1}{2}kt)-m-3\right)}{\pi \left(-4m^2\cos^4(\frac{1}{2}kt)+4m^2\cos^2(\frac{1}{2}kt)+4cm\cos(\frac{1}{2}kt)\sin(\frac{1}{2}kt)+c^2\right)}
\end{equation}

\begin{equation} \label{omega}
\omega(t)=-\frac{32mk^2(\pi+\frac{1}{16})\cos^2(\frac{1}{2}kt)-k^2(16\pi m+24\pi+m+3)}{k^2\left(2m\cos^2(\frac{1}{2}kt)-24\pi-m-3\right)}
\end{equation}

The cyclic behavior of pressure, energy density and cosmological constant versus cosmic time is shown in figure 1. The pressure is positive during the early-time where the expansion was decelerating, and negative during the late-time accelerating expansion without singular values or improper behavior. This pressure behavior is the same as in the standard cosmological model but singularity free. In the standard cosmological model \cite{decc} the early Universe ($z \rightarrow \infty$) is filled with positive pressure and the
expansion is decelerated, while in far future with DE domination the expansion is accelerated. The negative pressure represents a repulsive force that accelerates the expansion while the positive pressure dominates only during the decelerating expansion epoch. A very small amount of positive energy with negative pressure is supposed to push the universe to expand faster and faster. It has also been shown in \cite{posit} that a positive pressure with viscosity can lead to, in the framework of the causal Israel-Stewart formalism, a decelerated expansion. At the beginning of the cycle, the energy density starts from a large positive value and keeps decreasing to a very small positive value. It is worthy mentioning that while the cyclic evolution of $p$, $H$ and $a$ in the current $f(R,T)$ gravity model with negative $\Lambda(t)$ has no singular values, it possesses singularities in some other $f(R,T)$ gravity reconstructions. For example, using the time periodic deceleration parameter ansatz (\ref{q}) with the choice $f(R,T)=R+2\lambda T$ with no cosmological constant considered leads to singularities in the cyclic behavior of $p$, $\rho$, $H$ and $a$ \cite{sah}.   \par

Fig.1 (c) shows that the cosmological constant oscillates and keeps negative. The deceleration parameter is negative ($q<0$) for an accelerating universe and positive ($q>0$) for a decelerating universe. Fig.1(f) shows a signature flipping of the deceleration parameter, it starts as a positive decreasing function and keeps decreasing until it becomes negative then reverses back after finite time. So, for every single cycle the universe accelerates after an epoch of deceleration which agrees with recent observations. The deceleration parameter varies in the range $-3\lesssim q\lesssim 1$ through the whole cycle. The negative values of $q$ gives information about the general dynamical behavior: De Sitter expansion happens at $q=-1$, accelerating power-law expansion can be achieved for $-1 < q < 0$ and a super-exponential expansion happens for $q<-1$. Through the evolution of each cycle, the time varying EoS parameter $\omega(t)$ is restricted to the range $-2.25 \leq \omega(t) \lesssim \frac{1}{3}$ in a good agreement with observations \cite{ob1,ob2}. It starts as a decreasing function from a positive value $\simeq \frac{1}{3}$ (radiation-like era) and crosses the zero (dust era $\omega = 0$) to the negative domain. After it reaches the dark energy- dominated era (cosmological constant era) at $-1$, it crosses the phantom divide line at $\omega=-1$ to the phantom era ($\omega<-1$) showing a Quintom behavior. Quintom is a dynamical model of dark energy differs from cosmological constant, Quintessence, Phantom ...etc. by a promenint feature, namely its EoS can smoothly cross over $\omega=-1$. \par

The Quintom dynamics associated with $\omega=-1$ crossing gives rise to $\omega>-1$ in the past and $\omega<-1$ today which is supported by the observations \cite{quintom}. A bouncing model for a universe dominated by Quintom matter has been studied in \cite{quintom2}. For the second half of the cosmic cycle, $\omega$ is an increasing function starting from $-2.25$. Current observations indicate that constant $\omega$ is roughly equal to $-1$ which means that the accelerating universe could be in the cosmological constant era ($\omega=-1$), quintessence era ($-1<\omega<-1/3$) or phantom era ($\omega<-1$). Observations seem to favor dark energy with $\omega<-1$ \cite{vikman,obb1, obb2}, it has been argued that the dark energy rapidly evolving from the dust-like $\omega \simeq 0$ at high redshift $z \sim 1$ to phantom-like $-1.2 \lesssim \omega \lesssim -1$ at present $z \simeq 0$ provides the best fit for the supernova Ia data and CMBR measurements \cite{vikman}. The subject whether the dark energy can evolve to the phantom era or not has been studied by many authors. In \cite{vikman} it has been argued that If observations confirm the evolution of the dark energy dominating in the universe from $\omega \geq -1$ in the close past to $\omega <-1$ at the current epoch, then this transition can not be explained by the classical dynamics given by an effective scalar field Lagrangian. However, such transition is allowed in models which include more complicated physics \cite{d1,d2,d3}. \par 
Cosmological redshift is related to recession velocity $v$ through $z=\frac{v}{C}$ with $C$ the velocity of light. Using $a=\frac{1}{1+z}$, we get the following expression for the evolution of $\omega$ as a function of the redshift:

\begin{equation} \label{w}
\omega(z)=-\frac{f(z)}{k^2g(z)}
\end{equation}
where $f(z)$ and $g(z)$ are give by 
\begin{eqnarray*}
f(z)&=&\frac{32mk^2(\pi+\frac{1}{16})}{1+\frac{1}{c^2}\left(\tan \left(\frac{\sqrt{c^2-m^2}}{2}\ln\left(\frac{1}{A(1+z)}\right)\right)\sqrt{c^2-m^2}-m \right)^2}+k^2(-16\pi m -24 \pi-m-3)\\
g(z)&=&\frac{2m}{1+\frac{1}{c^2}\left(\tan \left(\frac{\sqrt{c^2-m^2}}{2}\ln\left(\frac{1}{A(1+z)}\right)\right)\sqrt{c^2-m^2}-m \right)^2}-24\pi-m -3
\end{eqnarray*}
For the current dark energy-dominated epoch, we have $z\simeq 0$ and $\omega=-1$. Therefore, we should pick the values for the constants $c$ and $m$ that gives $\omega=-1$ at $z=0$ and use them for all other parameters. We have noted that as the difference between $c$ and $m$ gets smaller, $\omega(z)$ approaches $-1$. This can be shown by calculating the limit of (\ref{w}), we find that
\begin{equation}
\lim_{c\rightarrow m} \omega(z)=-1
\end{equation}
Where we set $k=0.1$ and $A=1$. For example, setting $c=2.1$ and $m=2$ gives non-zero value for the scale factor (\ref{a}) at the beginning of time, $a(0)=51.34416870$ which means there is no initial singularity. Fig.1 (e) shows that the Hubble parameter $H$ oscillates and keeps positive which is not the case in bouncing cosmology where $H$ passes from $H < 0$ to $H > 0$ and vanishes at the bounce point.

\begin{figure}[H]
  \centering            
  \subfigure[$p$]{\label{F555}\includegraphics[width=0.3\textwidth]{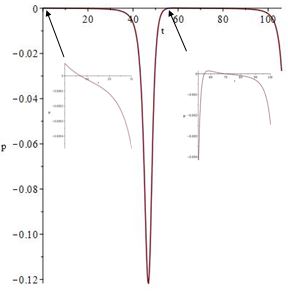}}
  \subfigure[$\rho$]{\label{F63}\includegraphics[width=0.3\textwidth]{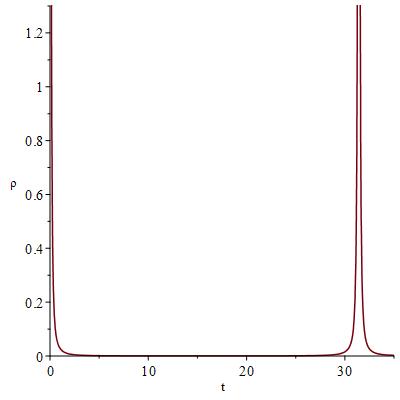}} 
	\subfigure[$\Lambda$]{\label{F423}\includegraphics[width=0.3\textwidth]{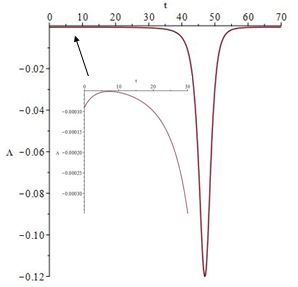}} \\
	 \subfigure[$\omega$]{\label{fig:rrf}\includegraphics[width=0.3\textwidth]{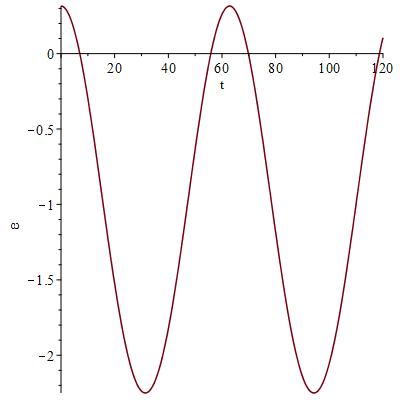}}                
  \subfigure[$H$]{\label{f222}\includegraphics[width=0.3\textwidth]{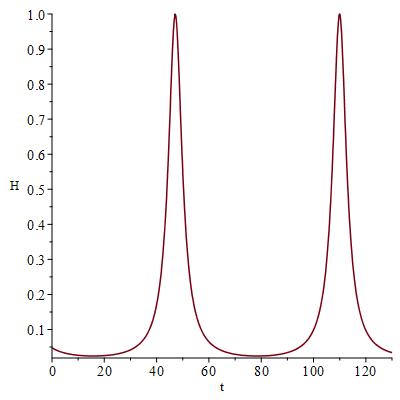}}
  \subfigure[$q(t)=m \cos kt -1$]{\label{F333}\includegraphics[width=0.3\textwidth]{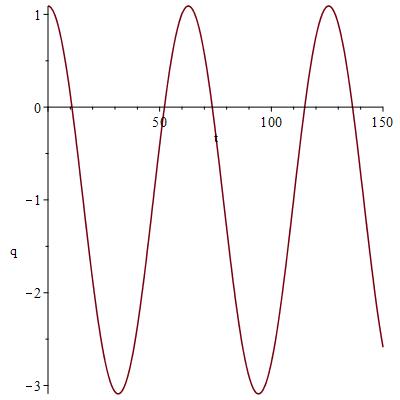}}\\ 
		\subfigure[$\rho-p$]{\label{F68}\includegraphics[width=0.3\textwidth]{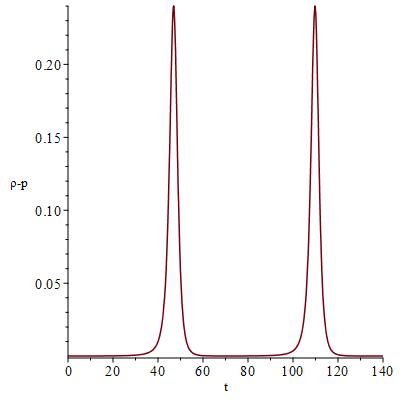}} 
	\subfigure[$\rho+p$]{\label{F680}\includegraphics[width=0.3\textwidth]{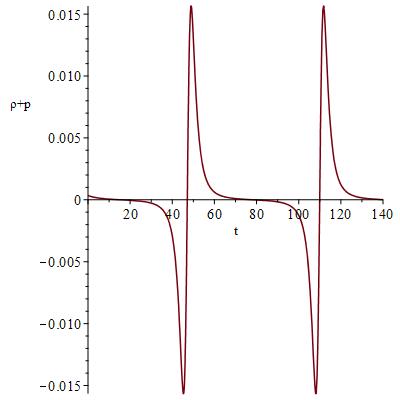}}
		\subfigure[$\rho+3p$]{\label{F67}\includegraphics[width=0.3\textwidth]{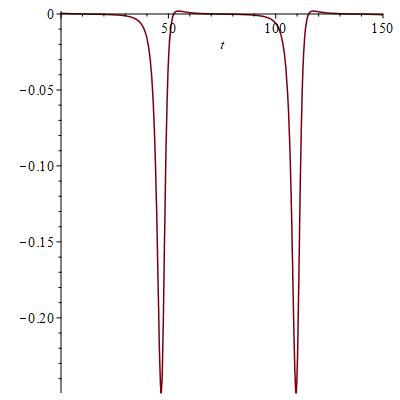}} 
  \caption{The behavior of $H$, $q$, $p$, $\rho$, $\Lambda$ and classical linear energy conditions verses cosmic time for $K=0$, $k=0.1$, $c=2.1$ and $m=2$. (a) The pressure is positive during the early-time and negative during the late-time without singular values. (b) The energy density starts from a large positive value and keeps decreasing to a very small positive value. (c) The cosmological constant is always negative through the whole cycle. (d) The EoS parameter $-2.25 \leq \omega(t) \lesssim \frac{1}{3}$ with $\omega=-1$ smooth crossing behavior. (e) The Hubble parameter oscillates but remains positive. (f) The deceleration parameter lies in the range $-3\lesssim q\lesssim 1$. (g) The DEC is valid all the time. (h) The WEC is valid for half cycle. (i) The SEC is valid only at the beginning of the cycle during the decelerating epoch where
attractive gravity dominates, and then becomes invalid during the late-time accelerating epoch dominated by repulsive gravity.}
  \label{fig:cassimir55}
\end{figure}

\section{Stability of solutions}

As we are ignoring the quantum effects in this model, the physical acceptability of the model can be checked through testing the classical linear energy conditions \cite{hawk,ec12} and the sound speed. The null, weak, strong and dominant energy conditions are respectively given by: $\rho + p \geq 0$; $\rho \geq 0$, $\rho + p \geq 0$; $\rho + 3p \geq 0$ and $\rho \geq \left|p\right|$. In the presence of semiclassical quantum effects, those classical linear energy conditions should be replaced by other nonlinear energy conditions \cite{ec}. It has been noted that these classical energy conditions are not fundamental physics and can not be valid in completely general situations \cite{ec2}. The strong energy condition (SEC) expresses the 'highly restrictive' statement that gravity should always be attractive. However, Even in the classical regime this condition fails when describing the universe in the current accelerated epoch and during inflation \cite{ec3,ec4,ec5}. For the current model, the negative pressure component represents a repulsive gravity and so we don't expect the SEC to be fulfilled in the late-time universe where this negative pressure component dominates. Fig.1 (i) shows that the SEC is valid only at the beginning of each cycle during the decelerating epoch where attractive gravity dominates, and then becomes invalid during the accelerating epoch where the repulsive gravity dominates. The dominant energy condition (DEC) expresses the fact that energy density should be non-negative and should propagate in a causal way. Fig.1 (g) shows that this condition is valid all the time. $\rho + p \geq 0$ is valid for the second half of the cycle. \par

The adiabatic square sound speed $c_s^2=\frac{dp}{d\rho}$ should be positive and less than $1$. Since causality implies that sound speed must be less than the speed of light, then the condition $0 \leq \frac{dp}{d\rho} \leq 1$ should be satisfied ($c=G=1$ in relativistic units). For the current model, we have
\begin{equation}
c_s^2=\frac{f_1(t)}{g_1(t)}
\end{equation}
where
\begin{eqnarray}
f_1(t)=-64m^2(\pi+\frac{1}{16}) \sin\left(\frac{1}{2}kt\right) \cos^5\left(\frac{1}{2}kt\right)+64m\left(m(\pi+\frac{1}{16})+\frac{3}{2}\pi+\frac{3}{16}\right)\times \\  \nonumber
\sin\left(\frac{1}{2}kt\right) \cos^3\left(\frac{1}{2}kt\right)+c(48\pi+6)\cos^2\left(\frac{1}{2}kt\right)-16\sin\left(\frac{1}{2}kt\right) \times\\ \nonumber 
\left(m^2(2\pi+\frac{1}{8})+m(3\pi+\frac{3}{8})+c^2(\pi+\frac{1}{16})\cos\left(\frac{1}{2}kt\right)\right)
-16c \left(m(\pi+\frac{1}{16})+\frac{3}{2}\pi+\frac{3}{16}\right).
\end{eqnarray}

\begin{eqnarray}
g_1(t)&=&4m^2 \sin\left(\frac{1}{2}kt\right) \cos^5\left(\frac{1}{2}kt\right)+4m(24\pi+m+3)\sin\left(\frac{1}{2}kt\right) \cos^3\left(\frac{1}{2}kt\right)\times \\  \nonumber
&-&c(48\pi+6)\cos^2\left(\frac{1}{2}kt\right)+(2m^2+m(48\pi+6)+c^2)\sin\left(\frac{1}{2}kt\right) \cos\left(\frac{1}{2}kt\right)  \\  \nonumber
&+&c(24\pi+m+3).
\end{eqnarray}
Figure (2) shows that the condition $0 \leq \frac{dp}{d\rho} \leq 1$ is satisfied only for a short period of time at the beginning of each cycle. 

\begin{figure}[H]
  \centering             
	\subfigure[$c_s^2$]{\label{F677}\includegraphics[width=0.3\textwidth]{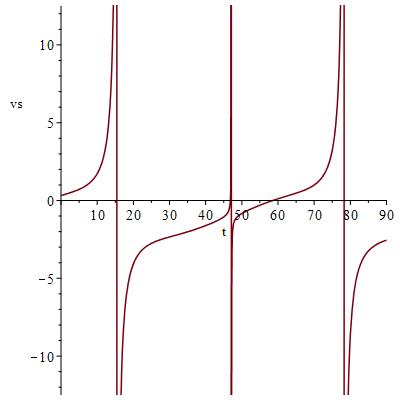}} 
  \caption{The behavior of $c_2^2$ verses cosmic time for $k=0.1$, $c=2.1$ and $m=2$.}
  \label{fig:cassimir552}
\end{figure}

\begin{table}[H]\label{tap}
\centering
\tiny
    \begin{tabular}{ | p{1.cm} | p{4cm} | p{4cm} | p{4cm} | }
    \hline
           & $K=0$ & $K=1$ & $K=-1$ \\ \hline
    p & -ve, but +ve for short period of time at the beginning of the cycle & -ve all the time  &  +ve all the time \\ \hline
    $\rho$ & +ve all the time & +ve all the time &  -ve all the time  \\ \hline
    $\Lambda$ & -ve all the time & -ve all the time & +ve all the time\\ \hline
		 $\omega$ & $-2.25 \leq \omega(t) \lesssim \frac{1}{3}$ & $-2.2 \leq \omega(t) \lesssim 0.2$ & $-2.3 \leq \omega(t) \lesssim 0.45$ with singularity at the end of the cycle  \\ \hline
		$\rho-p$ & valid all the time & valid all the time &  valid for short time at the beginning   \\ \hline
		$c_s^2$ & valid only at the beginning & valid only at the beginning & valid only at the beginning \\ \hline
		$p+\rho$ & valid for the second half of the cycle  & valid all the time & valid only at the beginning\\ \hline
		$\rho+3p$ & valid only at the beginning & valid only at the beginning  &  valid most of the time \\ \hline
    \end{tabular}
		\caption {The behavior of $p$, $\rho$, $\Lambda$, $c_s^2$ and energy conditions verses cosmic time for different $K$.}
		\end{table}
		
\section{Conclusion}

The current work explores a new oscillating cosmological scenario, in which the Hubble parameter oscillates and keeps positive, in a specific f(R,T) reconstruction with variable cosmological constant. The cosmological solutions have been obtained for the flat case supported by observations. This singularity-free model predicts a positive pressure during the early-time decelerated expansion, and a negative pressure during the late-time accelerating epoch. While the energy density and Hubble parameter both oscillate and keep positive, the time varying cosmological constant oscillates around a very small negative value and keeps negative. The evolution of the equation of state parameter shows a Quintom behavior and is restricted to the range $-2.25 \leq \omega(t) \lesssim \frac{1}{3}$. The existence of negative cosmological constant as a non-conventional mechanism that can address the late-time acceleration has been discussed. The stability of the model has been studied through testing the validity of the classical linear energy conditions and the sound speed causality condition. This different oscillating scenario in which the universe passes through the inflationary era periodically without singularities offers a good possibility to unify the late-time acceleration and inflation. The varying cosmological constant derived in the current work is nothing but the negative of the thermodynamical work density $W$. The suggested thermodynamical interpretation of $\Lambda$ in this paper has been compared to another thermodynamical interpretation of $\Lambda$ suggsted in the context of “Black hole chemistry”.

\section*{Acknowledgment}
We are so grateful to the reviewer for his valuable suggestions and comments that significantly improved the paper.

\end{document}